\documentclass[runningheads]{llncs}
\usepackage[T1]{fontenc}
\usepackage{graphicx}
\usepackage{amsmath,amssymb,amsfonts}
\usepackage{bm}
\usepackage{stmaryrd}
\usepackage{xspace}
\usepackage{cancel}
\usepackage{multirow}%
\usepackage{mathrsfs}%
\usepackage[title]{appendix}%
\usepackage{textcomp}%
\usepackage{manyfoot}%
\usepackage{booktabs}%
\usepackage{algorithm}%
\usepackage{algpseudocode}%
\usepackage{listings}%
\usepackage{cite}
\usepackage{lineno}
\usepackage{array}
\usepackage{xcolor}
\usepackage{bm}
\usepackage{hyperref}
\usepackage{tikz}
\usepackage{xcolor}
\usepackage{pifont}
\usetikzlibrary{positioning}
\usepackage{adjustbox}
\usepackage{siunitx}
\newcommand{\sem}[1]{\ensuremath{\llbracket #1 \rrbracket}\xspace}

\begin{document}
%
\title{A Pragmatic Guide to Building Conservative Discrete Abstractions of Cyber-Physical Systems}
\titlerunning{Building Conservative Abstractions of CPS}
%
\author{Jordan Peper \and Krish Kapadia \and James Gast \and Ethan Howes \and Ivan Ruchkin}
%
\authorrunning{J. Peper et al.}
%
\institute{University of Florida, Gainesville FL 32611, USA}
\maketitle              
\begin{abstract}
Symbolic model checking is an effective approach for verifying semantically rich temporal-logic properties of cyber-physical systems, but it hinges on discretizing continuous-state dynamics into a finite-state abstraction. To transfer verification guarantees from the abstract model to the concrete CPS, the abstraction must conservatively approximate the concrete state space and behaviors. Hence, model-builders must maintain this soundness while balancing pessimism with tractability. However, they face several common pitfalls such as under-approximating the state space, under-approximating transitions, unsound pruning of ``degenerate'' behaviors, and improper specification lifting. This tutorial presents a pragmatic, conservative-by-construction workflow for building discrete abstractions of closed-loop dynamical systems. The workflow consists of four modular steps with interchangeable subroutines: (i) state-space partition and abstraction-function design, (ii) conservative transition construction via axis-aligned bounding boxes, polytopes, or sampling with PAC coverage certificates, (iii) mitigation of spurious transitions and self-loops using certified erasure and counterexample-guided abstraction refinement, and (iv) sound lifting of LTL specifications using may-must semantics. We demonstrate the end-to-end pipeline on three case studies and report how these design choices affect abstraction structure, runtime, and verification outcomes.

\noindent \textbf{Code: } {\href{https://github.com/Trustworthy-Engineered-Autonomy-Lab/cps-abstraction-tutorial}{github.com/Trustworthy-Engineered-Autonomy-Lab} }\\
\end{abstract}

\section{Introduction}

Cyber-physical systems (CPS) operate in environments where failures threaten life, property, and the environment. Accordingly, there is sustained demand for rigorous correctness assurance for realistic CPS. Stemming from a long history of verification in hardware and software, symbolic model checking remains particularly effective for verifying CPS when requirements are expressed as semantically rich temporal-logic specifications~\cite{clarke_emerson_sistla_1986,baier_principles_2008}. Model checking performs an exhaustive, specification-driven exploration of a model's state space to determine whether a given safety property holds~\cite{baier_principles_2008, burch_clarke_mcmillan_1992}.

To be computationally tractable, model checking requires a \textit{finite-state} model, whereas real-world CPS evolve over continuous state spaces. Hence, engineers cast the CPS as a finite state model (called an \emph{abstraction} of the CPS) by aggregating the states into a finite set of abstract states, and then build transition relations over them. However, for guarantees to transfer from this abstraction to the concrete CPS, the abstraction must conservatively approximate the state space and behaviors of the concrete CPS.

\looseness=-1
In practice, creating a useful conservative abstraction requires carefully maintaining soundness and balancing pessimism against computational cost. Inexperienced model-builders tend to make four typical errors: (i) under-covering the concrete state space with the one, (ii) under-approximating the abstract transition relation (thereby omitting feasible behaviors), (iii) pruning degenerate behaviors without a certificate that preserves soundness, and (iv) lifting the specification without preserving the implication from abstract to concrete satisfaction.

This tutorial presents a single, sound \textit{model-building workflow} with standard techniques to avoid the aforementioned pitfalls. The workflow consists of four steps (Section~\ref{sec:step1}--\ref{sec:step4}) with interchangeable subroutines, resulting in a conservative-by-construction, sound abstraction. First, we explain how to construct an abstract state space using a rectilinear partition. Second, we present three approaches for building a conservative transition relation: (a) an axis-aligned bounding box over-approximation, (b) a less conservative polytope technique, and (c) a sampling-based construction equipped with a PAC-style certificate. Third, we describe common degenerate behaviors in conservative abstractions (including spurious transitions and self-loops) and mitigation strategies: certified self-loop erasure (with both a deterministic and probabilistic guarantee) and counterexample-guided abstraction refinement (CEGAR). Fourth and finally, we address the specification-lifting step by translating the concrete LTL formula to an abstract formula over the abstract state space to ensure that its satisfaction on the abstraction is sufficient to satisfaction on the concrete CPS.

The rest of this paper is organized as follows. Sec~\ref{sec:rw} revisits verification and abstraction literature for cyber-physical systems. Section~\ref{sec:pf} introduces relevant preliminaries used throughout the tutorial, including conservative abstraction, LTL lifting, and model checking. Section~\ref{sec:ovr} provides an in-depth look at the four abstraction steps discussed in detail in Section~\ref{sec:step1}--~\ref{sec:step4}. Finally, Section~\ref{sec:exp} demonstrates the full abstraction pipeline on three case studies, and then concludes with Section~\ref{sec:conc}.

\section{Literature Landscape}
\label{sec:rw}

\subsubsection{Verification of dynamical systems.} Verification of complex dynamical systems, including hybrid systems~\cite{henzinger_theory_1996}, is commonly organized into three methodological families: (i) deductive methods, (ii) explicit-state reachability analysis, and (iii) symbolic abstraction-based techniques~\cite{clarke_verification_2018, gueguen_safety_2009, alur_formal_2011}. Deductive methods exploit analytical system properties (e.g., invariants, Lyapunov arguments, or proof rules) to certify stability or safety~\cite{clarke_verification_2018}. Explicit-state reachability analysis propagates an initial set forward through the dynamics to compute an over-approximation of the reachable state tube, enabling detection of safety violations~\cite{chen_reachability_2022}. Finally, symbolic abstraction-based techniques construct a simpler finite-state model that soundly over-approximates the behaviors of the concrete system, thereby enabling the use of symbolic model checking~\cite{clarke_verification_2018}. This tutorial concentrates solely on abstraction-based techniques. 

\subsubsection{Abstraction of dynamical systems.} Casting a dynamical system as a symbolic abstraction requires reducing the state and action spaces to a finite-state over-approximation that preserves ground-truth behaviors. Typical abstraction schemes include phase-based~\cite{frehse_phaver_2008, frehse_spaceex_2011, sloth_algorithmic_2011}, where rectangular or linear flow constraints are placed on the system states and dynamics, and predicate-based~\cite{alur_counterexample-guided_2006, alur_predicate_2006, kloetzer_fully_2008}, where continuous state spaces are abstracted as a collection of predicate-defined subspaces. For abstracting the dynamics of a system, state-of-the-art tools include $\texttt{SCOTS}$~\cite{runggerSCOTSToolSynthesis2016}, which over-approximates reachable sets via a growth bound, and $\texttt{Flow}^*$~\cite{hutchison_flow_2013}, which leverages Taylor models to over-approximate reachable sets. When dynamics are not known, \cite{coppola_data-driven_2023, devonportSymbolicAbstractionsData2021a, lavaeiDataDrivenSynthesisSymbolic2022} sample trajectories from the system and synthesize probabilistic guarantees on model correctness.

\subsubsection{Counterexample-guided abstraction refinement (CEGAR).} Methodologies for mitigating the state explosion problem in symbolic model checking are rooted in early software verification, where automated refinement loops were developed to iteratively rewrite abstraction predicates until spurious abstract transitions are eliminated \cite{henzinger_lazy_2002}. A canonical representative is counterexample-guided abstraction refinement~\cite{clarke_counterexample-guided_2000}. Although these methods were initially developed for purely discrete-state systems, they were later extended to timed automata~\cite{dierks_automatic_2007, sloth_algorithmic_2011} and hybrid systems~\cite{goos_verification_2003, clarke_abstraction_2003, fehnker_refining_2005, alur_counterexample-guided_2006}. Subsequent work applied CEGAR to probabilistic abstractions, including Markov models~\cite{chadha_counterexample-guided_2010, chatterjee_cegar_2014, hermanns_probabilistic_2008}. This tutorial revisits CEGAR in the context of refining discrete abstractions. 

\section{Background}
\label{sec:pf}

\subsection{Systems}

We describe the preliminaries to conservative discrete-state abstraction, linear temporal logic (LTL) specification, and model checking. First, we assume that a discrete-time dynamical system is an infinite-state transition system.

\begin{definition} [Infinite-state transition system]
\label{def:ists}
    An infinite-state transition system is a tuple $s = (X, X_0, f)$, where:
    \begin{itemize}
        \item $X \subset \mathbb{R}^n$ is an $n$-dimensional state space
        \item $X_0 \subseteq X$ is a subset of initial states
        \item $f : X \rightarrow X$ is the dynamical transition function
    \end{itemize}
\end{definition}

\noindent
An infinite state \emph{trajectory} of this system is the sequence $\tau_s(x_0) = (x_0, x_1, x_2, \dots)$, where $x_{k+1} = f(x_k)$, and $x_0$ is any initial state in $X_0$. We aim to cast this dynamical system as a \emph{finite-state transition system}:

\begin{definition} [Finite-state transition system]
\label{def:fsts}
    A finite-state transition system is a tuple $\hat{s} = (\hat{X}, \hat{X}_0, \Sigma, L,\hat{f})$, where:
    \begin{itemize}
        \item $\hat{X} \subset \mathbb{N}^n$ is an $n$-dimensional abstract state set
        \item $\hat{X}_0 \subseteq \hat{X}$ is a subset of initial abstract states
        \item $\Sigma$ is a set of action labels
        \item $L : \hat{X} \rightarrow 2^\Sigma$ is the labeling function specifying the admissible actions at $\hat{x}$
        \item $\hat{f} : \hat{X} \times \Sigma \rightarrow \hat{X}$ is the transition function
    \end{itemize}
\end{definition}

\noindent
This finite-state transition system is \emph{nondeterministic}: each state $\hat{x}$ may admit multiple valid successors. These successors are indexed by the admissible labels $\sigma \in L(\hat{x})$ and are collected in the set $\mathrm{Post}_{\hat{s}}(\hat{x}) = \{\hat{f}(\hat{x}, \sigma) \mid \sigma \in L(\hat{x})\}$. A path of this system $\hat{\tau}$ initialized at $\hat{x}_0$ is an infinite sequence of states whose successive elements are valid successors. We denote the set of all paths from $\hat{x}_0$ by:
\begin{equation*}
    \mathrm{Paths}_{\hat{s}}(\hat{x}_0) = \{ (\hat{x}_0, \hat{x}_1, \dots) \mid \forall k \ge 0, \hat{x}_{k+1} \in \mathrm{Post}_{\hat{s}}(\hat{x}_k) \}.
\end{equation*}

\subsection{Abstraction}

\emph{Abstract} modeling is a process whereby an infinite-state transition (concrete) system is modeled as a finite-state transition (abstract) system whose states represent subsets of the concrete state space and transitions reflect those of the concrete dynamics. \emph{State abstraction} is performed by a \emph{state quantization} function $\psi : X \rightarrow \hat{X}$. For any $\hat{x} \in \hat{X}$, define the quantization cell (preimage) as $\Psi(\hat{x}) = \{x\in X \mid \psi(x) = \hat{x} \}$.

After the state space is abstracted into a finite collection of abstract states, the abstract dynamics are built over the abstract states. If the resulting finite-state transition system contains all concrete states, initial states, and one-step transitions under this quantization, we call it a \emph{conservative abstract model}~\cite{goos_verification_2003}.

\begin{definition} [Conservative abstract model]
\label{def:cons-abstraction}
    Let $s$ be an infinite-state transition system (Definition~\ref{def:ists}) and let $\hat{s}$ be a finite-state transition system (Definition~\ref{def:fsts}) related to $s$ through the a state quantization function $\Psi$. The system $\hat{s}$ is a \emph{conservative abstraction} of $s$ if the following conditions hold:
    \begin{enumerate}
        \item $X \subseteq \bigcup_{\hat{x} \in \hat{X}} \Psi(\hat{x})$ and $X_0 \subseteq \bigcup_{\hat{x}_0 \in \hat{X}_0} \Psi(\hat{x}_0)$
        \item $\forall \hat{x} \in \hat{X},f(\Psi(\hat{x})) \subseteq \bigcup_{\hat{x}'\in \mathrm{Post}_{\hat{s}}(\hat{x})} \Psi(\hat{x}')$
    \end{enumerate}
\end{definition}
Condition 1 that every concrete state and initial state have one abstract representative. Condition 3 ensures that every one-step transition of the concrete system is represented by a transition in the abstract model.

\subsection{Conservative LTL lifting}
We assume that we are given a safety specification $\varphi$ of the concrete system $s$, expressed in linear temporal logic (LTL):

\begin{definition} [Linear temporal logic specification]
\label{def:ltl}
    A linear temporal logic specification is logical formula built from the syntax:
    \[
    \varphi ::= \Box \phi \mid \Diamond \phi \mid \mathcal{U} \phi \mid \bigcirc \phi \mid \phi_1 \wedge \phi_2 \mid \phi_1 \vee \phi_2 \mid \neg \phi \mid p(x) \mid \top \mid \bot
    \]
    \begin{itemize}
        \item $\Box$, $\Diamond$, $\mathcal{U}$, and $\bigcirc$ denote the globally, eventually, until, and next operators
        \item $\wedge$ and $\vee$ denote the logical AND and OR operators 
        \item $\neg$ denotes logical negation 
        \item $p : X \rightarrow \{\top, \bot \}$ is an atomic predicate on the state $x \in X$ (e.g., $|x| \ge 3$)
        \item $\top$ and $\bot$ denote the logical TRUE and FALSE constants
    \end{itemize}
\end{definition}
The satisfaction of an LTL specification on a trajectory of the concrete system is denoted $\tau_s(x_0) \vDash \varphi$. Similarly, if $\tau_s(x_0) \vDash \varphi$ for all $x_0 \in X_0$, then the system satisfies the specification, denoted simply as $s \vDash \varphi$.

This property is expressed through atomic propositions on the domain $X$. However, we aim to verify that this property holds on the conservative abstraction $\hat{s}$, which possesses the coarser abstract state space $\hat{X}$. Hence, we must build a \emph{lifted} specification $\hat{\varphi}$ from $\varphi$. The lifted specification $\hat{\varphi}$
holds on the trajectories of $\hat{s}$ stemming from $\hat{x}_0$ ($\mathrm{Paths}_{\hat{s}}(\hat{x}_0)$) if and only if the specification holds on every trajectory in $\mathrm{Paths}_{\hat{s}}(\hat{x}_0)$, that is:
\begin{equation}
\label{eqn:abs-sat}
    \mathrm{Paths}_{\hat{s}}(\hat{x}_0) \vDash \hat{\varphi} \iff \forall \hat{\tau} \in \mathrm{Paths}_{\hat{s}}(\hat{x}_0), \hat{\tau} \vDash \hat{\varphi}.
\end{equation}
To ensure that the verification guarantees on $\hat{s}$ transfer to the concrete system $s$, we must ensure that $\hat{\varphi}$ is a \emph{conservative LTL translation} of $\varphi$:

\begin{definition} [Lifted LTL specification]
\label{def:cons-trans}
    Given a system $s$, an LTL safety property $\varphi$, and a conservative abstraction $\hat{s}$ induced by $\psi$, the LTL specification $\hat{\varphi}$ is conservatively lifted to $\hat{s}$ if, for any $\hat{x}_0 \in \hat{X}_0$:
    \begin{equation}
       \forall x_0 \in \Psi(\hat{x}_0): \big ( \mathrm{Paths}_s(\hat{x}_0) \vDash \hat{\varphi} \implies \tau_s(x_0) \vDash \varphi  \big ).
    \end{equation}
    Equivalently, $\hat{s} \vDash \hat{\varphi} \implies s \vDash \varphi$.
\end{definition}
This tutorial covers conservative lifting of $\varphi$ to the system abstraction $\hat{s}$. 

\subsection{Conservative Model Checking}
Model checking performs an exhaustive, specification-driven exploration of a model's state space to determine whether that specification holds~\cite{baier_principles_2008, burch_clarke_mcmillan_1992}. To model check the \emph{conservative system abstraction} $\hat{s}$ from an initial abstract state $\hat{x}_0$, an algorithm explores all abstract trajectories $\hat{\tau} \in \mathrm{Paths}_s(\hat{x}_0)$ that originate from $\hat{x}_0$ under the nondeterminism of the abstraction.

Equation~\ref{eqn:abs-sat} establishes an important conservatism property: the lifted safety specification $\hat{\varphi}$ is satisfied from $\hat{x}_0$ if and only if $\hat{\tau} \vDash \hat{\varphi}$ for all $\hat{\tau} \in \mathrm{Paths}_s(\hat{x}_0)$. If there exists any trajectory in $\mathrm{Paths}_s(\hat{x}_0)$ that violates $\hat{\varphi}$, then the abstraction is not verified from $\hat{x}_0$; that is, $\mathrm{Paths}_s(\hat{x}_0) \nvDash \hat{\varphi}$.However, if we find that $\mathrm{Paths}_s(\hat{x}_0) \vDash \hat{\varphi}$, then for all $x \in \Psi(\hat{x}_0)$, $\tau_s(x) \vDash \varphi$. Consequently, it is impossible for the abstraction to verify the property while the concrete system violates it.

\begin{figure}[h]
    \centering
    \includegraphics[width=1.0\textwidth]{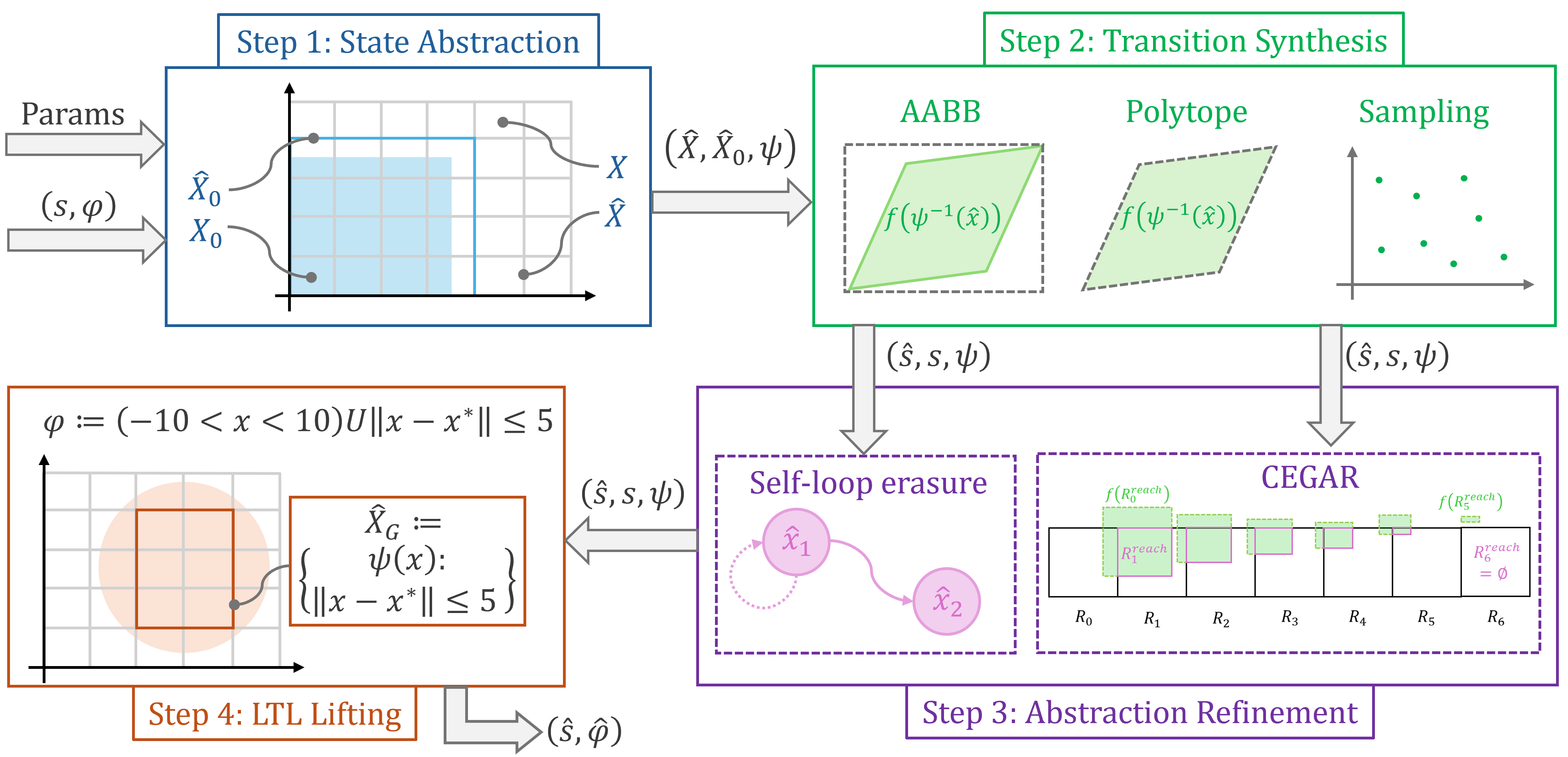}
    \caption{Overview of the four-step abstraction workflow.}
    \label{fig:ovr}
\end{figure}

\section{Overview of Conservative System Abstraction}
\label{sec:ovr}

We start with a brief overview of the conservative system abstraction pipeline (illustrated in Figure~\ref{fig:ovr}), followed by a running example that we use to exemplify each stage. The abstraction pipeline takes two inputs: (i) a concrete dynamical model (Definition~\ref{def:ists}) and (ii) an LTL safety specification (Definition~\ref{def:ltl}). From these inputs, conservative abstraction proceeds through four steps: (1) abstract the state space, (2) build a conservative transition relation, (3) diagnose and remove spurious behaviors via refinement, and (4) lift the LTL specification to the abstract model. The output is a conservative abstraction of the system in the form of a \emph{finite state machine} (Definition~\ref{def:fsts}) and a lifted specification used in symbolic model checking.

In Step~1 (Section~\ref{sec:step1}), we define a \emph{quantization function} $\psi$ from user-selected hyperparameters (e.g., the coarseness along each dimension), then construct the abstract state set $\hat{X}$ and abstract initial set $\hat{X}_0$ and ensure that Conditions~(1) and (2) of Definition~\ref{def:cons-abstraction} are met. In Step~2 (Section~\ref{sec:step2}), we build the abstract transition relation $\hat{f}$ over $\hat{X}$ so that Condition~(3) of Definition~\ref{def:cons-abstraction} holds, and we compare several approaches suited to different classes of dynamics $f$ (affine, nonlinear, black-box). In Step~3 (Section~\ref{sec:step4}), we identify spurious behaviors (self-loops in $\hat{f}$ and spurious counterexamples) and refine both the state space $\hat{X}$ and transition relation $\hat{f}$ to eliminate these behaviors without sacrificing soundness. Finally, in Step~4 (Section~\ref{sec:step3}), we lift the LTL specification from $X$ to the refined abstract state space $\hat{X}$, ensuring the condition in Definition~\ref{def:cons-trans} holds.

\subsubsection{Running example.} We will use a unicycle model as a running example throughout this tutorial. The unicycle state is given by $x = (y_1, y_2, \theta) \in [0.0, 50.0] \times [0.0, 40.0] \times [-\pi, \pi]$, where $y_1$ and $y_2$ represent the two positional coordinates, and $\theta$ denotes the system's heading angle. The objective of this system is to reach within $8.0$ units of the goal located at $(y_1^{goal}, y_2^{goal}) = (40.0, 20.0)$ while remaining in bounds and at least $5.0$ from an obstacle centered at $(y_1^{obs}, y_2^{obs}) = (25.0, 25.0)$. This specification expressed in LTL:
\begin{align}
\label{eqn:run-ltl}
   & \varphi := [(0 < y_1 < 50)\wedge (0 < y_2 < 40)\wedge d_1(y_1, y_2) > 5] ~\mathcal{U}~(d_2(y_1, y_2) \le 8); \\
    & d_1(y_1, y_2) := \sqrt{(y_1-25)^2+(y_2-25)^2}; \quad d_2(y_1, y_2) := \sqrt{(y_1-40)^2+(y_2-20)^2} \notag
\end{align}
The system dynamics are given by the Dubins model~\cite{dubins_curves_1957}:
\begin{equation}
\label{eqn:run-dyn}
    f(x) = 
    \begin{bmatrix}
        y_1[k+1] \\
        y_2[k+1] \\
        \theta[k+1]
    \end{bmatrix}
    =
    \begin{bmatrix}
        y_1[k] \\
        y_2[k] \\
        \theta[k]
    \end{bmatrix}
    +
    \begin{bmatrix}
        v\cos(\theta[k]) \\
         v\sin(\theta[k]) \\
         u[k]
    \end{bmatrix}
    \Delta t,
\end{equation}
where $\Delta t = 0.5$ is a fixed sampling time, $v = 5.0$ is the constant unicycle speed, and $u$ is the control action (heading rate of change) selected from a deterministic state controller designed for goal-attraction and obstacle-repulsion (see the Appendix).

\section{Step 1: Abstract the State-Space}
\label{sec:step1}

\subsubsection{State-space quanitzation.} A finite-state transition system (Definition~\ref{def:fsts}) requires a \textit{finite} collection of states to evolve over. However, the concrete model does not fit this template due to the continuous domain $X$. One solution is to formulate a \emph{state-space abstraction} by imposing a uniform rectilinear grid over $X$ whose induced subsets compose the abstract states $\hat{X}$. Suppose $X$ is a rectangular domain $[ \underline{x}^{(1)}, \overline{x}^{(1)}]  \times \dots \times [ \underline{x}^{(n)}, \overline{x}^{(n)}]$, where $\underline{x}^{(j)}$ and $\overline{x}^{(j)}$ denote the boundaries of the domain along dimension $j \in 1, \dots, n$. Let $c^{(j)}$ be a user-defined count of desired abstract states per dimension $j$. Then we specify the set of uniformly-spaced breakpoints along each dimension $\mathcal{B} = \{ B^{(1)}, \dots, B^{(n)} \}$, where:
\begin{equation*}
    B^{(j)} = \left( \underline{x}^{(j)} < b^{(j)}_2 < \dots < b^{(j)}_{c^{(j)}} < \overline{x}^{(j)} \right), \quad b^{(j)}_2 - \underline{x}^{(j)} =  \dots = \overline{x}^{(j)} - b^{(j)}_{c^{(j)}},
\end{equation*}
is the tuple of uniformly spaced breakpoints along the $j^{\text{th}}$ dimension, including the boundaries. Let $\text{i} = (i_1, \dots, i_n) \in I$ be a vector of indices where $I = \{1, \dots, c^{(1)}\} \times \dots \times \{1, \dots, c^{(n)}\}$ is the valid index set. The induced abstract state set is the collection of abstract states $\hat{x}[\text{i}]$ for all possible indices $\text{i}$, that is, $\hat{X} = \{ \hat{x}[\text{i}] \mid \text{i} \in I \} = \{ \psi(x) \mid x \in X \}$. The initial abstract state set is also given by $\hat{X}_0 = \{\psi(x_0) \mid x_0 \in X_0  \}$, where the abstraction function $\psi$ is defined as
\begin{equation}
    x \in \prod_{j=1}^n \left [b_{i_{j}}^{(j)}, b_{i_{j} + 1}^{(j)} \right ] \implies \psi(x) = \hat{x}[\text{i}],
\end{equation}
and its inverse is defined as
\begin{equation}
\label{eqn:conc}
    \Psi(\hat{x}[\text{i}]) = \{ x \in \mathbb{R}^n : b_{i_{j}}^{(j)} \le x^{(j)} < b_{i_{j} + 1}^{(j)}, \,\,\, j = 1, \dots, n \}.
\end{equation}
Both $\hat{X}$ and $\hat{X}_0$ satisfy conditions~(1)–(2) of Definition~\ref{def:cons-abstraction}. 


\subsubsection{Running example.} Recall the state space of the unicycle system $X = [0.0, 50.0] \times [0.0, 40.0] \times [-\pi, \pi]$. To match the above notation, we represent the state vector as $x = (y_1, y_2, \theta) = (x^{(1)}, x^{(2)}, x^{(3)})$. Suppose we choose $c^{(1)} = c^{(2)}=c^{(3)} = 10$ to be the desired number of abstract state cells along each dimension ($10^3$ total abstract states). The set of breakpoints is $\mathcal{B} = \{ B^{(1)}, B^{(2)}, B^{(3)} \}$, where:
\begin{equation*}
    B^{(1)} = \{0, 4, \dots, 40\}, \quad B^{(2)} = \{0, 5, \dots, 50\}, \quad B^{(3)} = \left\{-\pi, -\frac{4\pi}{5}, \dots, \pi \right\}
\end{equation*}
In this example, the index $\text{i}$ of any abstract state is given by $\text{i} = (i_1, i_2, i_3 )$ where $i_1, i_2, i_3, \in 1, \dots, 10$. Suppose we want to concretize the abstract state $\hat{x}[3, 2, 1]$. Following Equation~\ref{eqn:conc}, we get the subspace:
\begin{align*}
    \Psi(\hat{x}[3, 2, 1]) = \{ x \in \mathbb{R}^n : b_{3}^{(1)} < x^{(1)} < b_{4}^{(1)}, b_{2}^{(2)} < x^{(2)} < b_{3}^{(2)}, b_{1}^{(3)} < x^{(3)} < b_{2}^{(3)} \} \\
    =\{ x \in \mathbb{R}^n : 8 < x^{(1)} < 12, 5 < x^{(2)} < 10, -\pi < x^{(3)} <4\pi/5 \} \subset X
\end{align*}

\section{Step 2: Build the Conservative Transitions}
\label{sec:step2}

With the abstract state space $\hat{X}$ in hand, the next step is to construct the transition relation $\hat{f}$ and the labeling function $L$. This construction follows a simple deterministic loop over $\hat{X}$ described in Algorithm~\ref{alg:abs-sys}:
\begin{equation*}
    (L,\hat{f}) = \texttt{BuildTransitions}(\psi,\hat{X},f)
\end{equation*}
The only nontrivial component of this routine is the procedure \texttt{GetSuccessors}, which computes, for each $\hat{x} \in \hat{X}$, a set of successor abstract states $\mathrm{Post}_{\hat{s}}(\hat{x})$. This section presents two interchangeable implementations of \texttt{GetSuccessors}: (1) a conservative axis-aligned bounding-box (AABB) method (Algorithm~\ref{alg:aabb}) and (2) a tighter polytope-based method (Algorithm~\ref{alg:conv}). Then, we present a sample-based replacement of \texttt{BuildTransitions}, which yields a statistical, not exhaustive, guarantee on the validity of the induced transitions (Algorithm~\ref{alg:sample}).

\begin{algorithm}[h]
\caption{\texttt{BuildTransitions}: build the conservative transition relation $\hat{f}$}
\label{alg:abs-sys}
\begin{algorithmic}[1]
\Require abstraction function $\psi : X \rightarrow \hat{X}$; abstract state space $\hat{X}$; dynamics $f$
\Ensure labeling function $L \subseteq \hat{X} \times 2^{\Sigma}$; transition relation $\hat{f} \subseteq \hat{X} \times \Sigma \times \hat{X}$

\State $L, \hat{f} \gets \emptyset$, $\Sigma \gets \{\sigma_{11},\sigma_{12},\dots, \sigma_{ij}, \dots\}$ 

\ForAll{$\hat{x}_i \in \hat{X}$}
    \State $\mathrm{Post}_{\hat{s}}(\hat{x}) \gets \texttt{GetSuccessors}(\hat{x}_i,\psi,\hat{X},f)$
    \ForAll{$\hat{x}_j \in \mathrm{Post}_{\hat{s}}$}
        \State $L \gets L \cup \{(\hat{x}_i, \sigma_{ij})\}$
        \Comment{$\sigma_{ij}$ is unique to the ordered pair $(\hat{x}_i,\hat{x}_j)$}
        \State $\hat{f} \gets \hat{f} \cup \{(\hat{x}_i,\sigma_{ij},\hat{x}_j)\}$
    \EndFor
\EndFor

\State \Return $L, \hat{f}$
\end{algorithmic}
\end{algorithm}

\subsubsection{AABB propagation.} We first present a conservative, AABB-based technique for computing the successor set $\mathrm{Post}_{\hat{s}}(\hat{x}) \subset \hat{X}$ for any $\hat{x} \in \hat{X}$ according to $\psi$ and $f$. Let $R \subset X$ be a region in the $n$-dimensional state space. An \emph{axis-aligned bounding box} of $R$ is a rectilinear subset of $X$ that inscribes $R$:
\begin{equation}
\label{eqn:bbox}
    \operatorname{AABB}(R) := \prod_{j = 1}^n [l^{(j)}, u^{(j)}], \quad l_j = \min_{x \in R}x^{(j)}, \quad u_j = \max_{x \in R}x^{(j)},
\end{equation}
where $l^{(j)}$ and $u^{(j)}$ are the states in $R$ containing the min/max $j$th component.

To conservatively approximate the successors of $\hat{x}$, we compute the AABB of its concrete image, and then determine the set of states that this image over-approximation intersects with. Let $f(\Psi(\hat{x}))$ be the concrete image of $\hat{x}$ under $f$. Then, its AABB over-approximation is denoted $\operatorname{AABB}(f(\Psi(\hat{x})))$. To build the successor set, Algorithm~\ref{alg:aabb} iterates over all candidate successors $\hat{x} \in \hat{X}$ and determines whether their concretization intersects with this bounding box. If so, this is allocated as a transition relation in $\hat{f}$. This technique, paired with Algorithm~\ref{alg:abs-sys}, yields a transition relation that satisfies condition~(3) of Definition~\ref{def:cons-abstraction}.

\begin{algorithm}[h]
\caption{\texttt{GetSuccessorsAABB}: compute successors of $\hat{x}$ according to the AABB image over-approximation}
\label{alg:aabb}
\begin{algorithmic}[1]
\Require abstract state $\hat{x}$; abstraction function $\psi : X \rightarrow \hat{X}$; abstract state space $\hat{X}$; dynamics $f$
\Ensure successor set $\mathrm{Post}_{\hat{s}} \subseteq \hat{X}$

\State $ \mathrm{Post}_{\hat{s}} \gets \emptyset$, $R \gets \Psi(\hat{x})$
\State $R' \gets \operatorname{AABB}(f(R))$ \Comment{Over-approximate image of $R$ via AABB (Eqn.~\ref{eqn:bbox})}


\ForAll{$\hat{x}' \in \hat{X}$} \Comment{Iterate over candidate successors}
    \If{$\Psi(\hat{x}') \cap R'$}
        \State $\mathrm{Post}_{\hat{s}} \gets \mathrm{Post}_{\hat{s}} \cap \{\hat{x}'\}$
    \EndIf
\EndFor

\State \Return $ \mathrm{Post}_{\hat{s}}$
\end{algorithmic}
\end{algorithm}

When the concrete system dynamics $f$ are linear in $X$, the minimizing and maximizing arguments in Equation~\ref{eqn:bbox} of $\operatorname{AABB}(f(R))$ are simply the minimizing and maximizing arguments of the propagated set of corners $\{f(l^{(1)}), f(u^{(1)})\} \times \dots \times \{f(l^{(n)}), f(u^{(n)})\}$. When $f$ is nonlinear but at least differentiable in $X$, the literature presents several techniques to determine an AABB over-approximation of $f(R)$. In the popular abstraction $\texttt{SCOTS}$~\cite{runggerSCOTSToolSynthesis2016}, the centroid of $R$ is propagated forward and a conservative growth bound based on the Lipschitz constant of $f$ is used to build an AABB over-approximation of $f(R)$. Another technique used in the popular reachability analysis tool $\texttt{FLOW}^*$~\cite{hutchison_flow_2013} builds a first-order Taylor approximation $F \approx f$, propagates the corners of $R$ through $\hat{f}$, and then inflates the min and max vertices in this set by the first-order Taylor remainder.

\subsubsection{Polytope propagation.} AABBs over-approximate the image of the concretized abstract state relatively tightly when $f$ is linear. However, in the nonlinear setting, the mapping from a concretized abstract state to its image through $f$ may involve shear, stretching, and even folding. Hence, an AABB may be too conservative and lead to spurious edges in $\hat{f}$. A less conservative technique is to over-approximate the image with a polytope $\operatorname{PT}$:
\begin{equation}
\label{eqn:poly}
     \operatorname{PT}(R) := \left\{ \sum_{k=1}^{2^n} \lambda_k \mid x \in R, \lambda_k \ge 0, \sum_{k=1}^{2^n} \lambda_k = 1 \right\},
\end{equation}
where $\lambda_k$ are convex-combination coefficients used to form the convex hull over $R$. We denote the polytope-based image over-approximation of $\hat{x}$ as $\operatorname{PT}(f(\Psi(\hat{x})))$.

To build the successor set using the polytope over-approximation of $f(\Psi(\hat{x}))$, Alg.~\ref{alg:conv} first rules out obvious non-intersecting subsets $\Psi(\hat{x})$ using the bounding box over-approximation $\operatorname{AABB}(f(\Psi(\hat{x})))$ (which inscribes the polytope), and then iterates over the remaining states to build the successor set. This greatly improves the performance of the algorithm since AABB intersection tests are faster than the polytope tests. Like the AABB method, this technique yields a transition relation that satisfies condition~(3) of Definition~\ref{def:cons-abstraction}.

\begin{algorithm}[h]
\caption{\texttt{GetSuccessorsPT}: compute successors of $\hat{x}$ using polytope over-approximation with AABB culling}
\label{alg:conv}
\begin{algorithmic}[1]

\Require abstract state $\hat{x}$; abstraction function $\psi : X \to \hat{X}$; abstract state space $\hat{X}$; dynamics $f$; 
\Ensure successor set $\mathrm{Post}_{\hat{s}} \subseteq \hat{X}$

\State $\mathrm{Post}_{\hat{s}} \gets \emptyset$, $R \gets \Psi(\hat{x})$
\State $R_{\text{AABB}}' \gets \operatorname{AABB}(f(R))$ \Comment{Over-approximate image of $R$ via AABB (Eqn.~\ref{eqn:bbox})}
\State $R_{\text{PT}}' \gets \operatorname{PT}(f(R))$ \Comment{Over-approximate image of $R$ via PT (Eqn.~\ref{eqn:poly})}


\State $\mathcal{C} \gets \emptyset$ \Comment{Define set of filtered candidates}
\ForAll{$\hat{x}' \in \hat{X}$}
    \If{$\Psi(\hat{x}') \cap R_{\text{AABB}}'$}
        \State $\mathcal{C} \gets \mathcal{C} \cup \{\hat{x}'\}$ \Comment{Allocate as a potential candidate}
    \EndIf
\EndFor

\ForAll{$\hat{x}' \in \mathcal{C}$}
    \If{$\Psi(\hat{x}') \cap R_{\text{PT}}'$}
        \State $\mathrm{Post}_{\hat{s}} \gets \mathrm{Post}_{\hat{s}} \cup \{\hat{x}'\}$
    \EndIf
\EndFor

\State \Return $ \mathrm{Post}_{\hat{s}}$

\end{algorithmic}
\end{algorithm}
\vspace{-6mm}

\subsubsection{Sampling-based successors.} When the dynamics $f$ are strongly nonlinear, black-box, or the rectilinear partition is coarse, sampling can yield the least conservative approximation of the successor set $\mathrm{Post}_{\hat{s}}(\hat{x})$. The trade-off is that, unlike the above propagation methods, sampling may fail to observe a genuine transition,  feasible in the concrete system. Consequently, the transition relation $\hat{f}$ produced by sampling does not deterministically satisfy Definition~\ref{def:cons-abstraction}. 

Instead, one can make a probabilistic guarantee of the following form: after $N$ samples, $\hat{f}$ contains all ground-truth transitions with confidence at least $1-\alpha$. To make this precise, we first describe a ``ground-truth'' guarantee expressed in terms of the true transition relation (which we denote $\hat{f}^*$) and then use it as a template for the data-driven setting where $\hat{f}^*$ is unknown and only sampled observations of $f$ are available. Algorithm~\ref{alg:sample} describes this data-driven method for building and certifying $\hat{f}$, replacing Algorithm~\ref{alg:abs-sys}.

\begin{algorithm}[h]
\caption{\texttt{SampleTransitions}: build $\hat{f}$ by sampling from $\operatorname{Unif}(X)$ until the certificate condition (Equation~\ref{eqn:cert}) is met}
\label{alg:sample}
\begin{algorithmic}[1]
\Require abstraction function $\psi : X \rightarrow \hat{X}$; abstract state space $\hat{X}$; state space $X$; dynamics $f$; confidence level $\delta$; significance level $\beta$
\Ensure labeling function $L \subseteq \hat{X} \times 2^{\Sigma}$; transition relation $\hat{f} \subseteq \hat{X} \times \Sigma \times \hat{X}$

\State $L \gets \emptyset$, $\hat{f} \gets \emptyset$, $\Sigma \gets \{\sigma_{11},\sigma_{12},\dots, \sigma_{ij}, \dots\}$
\State $\overline{M}_N \gets 1.0$, $N \gets 0$, $c_1 \gets 0$
\State $C[\cdot] \gets \emptyset$ \Comment{Map: triple $\mapsto$ transition allocation count}

\While{$\overline{M}_N \ge \beta$}
    \State $x \gets \operatorname{Unif}(X)$, $\hat{x}_i \gets \psi(x)$, $\hat{x}_j \gets \psi(f(x))$
    \State $t \gets (\hat{x}_i,\sigma_{ij},\hat{x}_j)$ \Comment{$\sigma_{ij}$ unique to ordered pair $(\hat{x}_i,\hat{x}_j)$}

    \If{$t \notin \mathrm{dom}(C)$}
        \State $C[t] \gets 1$, $c_1 \gets c_1 + 1$
        \State $L\gets L \cup \{(\hat{x}_i, \sigma_{ij})\}$
        \State $\hat{f} \gets \hat{f} \cup \{t\}$
    \ElsIf{$C[t] = 1$}
        \State $C[t] \gets 2$, $c_1 \gets c_1 - 1$
    \Else
        \State $C[t] \gets C[t] + 1$
    \EndIf

    \State $N \gets N + 1$
    \State $\overline{M}_N \gets \frac{c_1}{N} + (2\sqrt{2} + \sqrt{3})\sqrt{\frac{\ln(3/\delta)}{N}}$ \Comment{(Equation~\ref{eqn:miss-mass})}
\EndWhile
\State \Return $\hat{f}$, $N$, $\overline{M}_N$

\end{algorithmic}
\end{algorithm}
\vspace{-4mm}

\paragraph{Ground-truth guarantee.} Consider a uniform distribution over the state space $X$, denoted $\operatorname{Unif}(X)$. This induces a categorical distribution $\operatorname{Cat}(\hat{f}^*)$ over transition relations $(\psi(x), \sigma, \psi(f(x)))$ in $\hat{f}^*$ with categories $\ell =  \{1, \dots,  K\}$, where $K = |\hat{f}^*|$ is the number of ground-truth transitions in $\hat{f}^*$. Each triple $(\psi(x), \sigma, \psi(f(x)))$ with index $\ell$ has a probability mass $q_{\ell}$ in this $\operatorname{Cat}(\hat{f}^*)$. Intuitively, $q_{\ell}$ is the chance that after a single draw $x \sim \operatorname{Unif}(X)$, the algorithm allocates the relation $(\psi(x), \sigma, \psi(f(x)))$ to $\hat{f}$. The exact formula for $q_{\ell}$ can be found in the Appendix.

We are ultimately interested in the probability that $N$ i.i.d. draws suffice to recover \emph{all} true transitions in $\hat{f}^*$ (precisely, $\hat{f} = \hat{f}^*$). By the inclusion-exclusion principle, this probability is given by:
\begin{equation}
    \operatorname{Pr}(\hat{f}^* = \hat{f})
    = \sum_{L \subseteq \{1, \dots, K\}} (-1)^{|L|} \left( 1-\sum_{\ell \in L} q_{\ell} \right)^{N} = 1 - \alpha,
\end{equation}
where $L$ is a subset of the category labels. Finally, observing all transitions implies that every concrete one-step behavior is represented in the abstract relation, so $\hat{f}^* = \hat{f}$ coincides with satisfaction of Condition~(3) in Definition~\ref{def:cons-abstraction}. Hence, Condition~(3) is satisfied with confidence $1-\alpha$.

\paragraph{Probably-approximately correct (PAC) guarantee.} In practice, we do not have access to the parameters of the induced categorical distribution $\operatorname{Cat}(\hat{f}^*)$ ($q_\ell$ and even the support size $K$ are unknown). Nevertheless, we can still sample transition relations $(\psi(x), \sigma, \psi(f(x)))$ from this categorical distribution $\operatorname{Cat}(\hat{f}^*)$ by sampling from $\operatorname{Unif}(X)$.

Let $\hat{f}_N \subseteq \hat{f}^*$ denote the set of relations discovered after drawing $N$ samples from $\operatorname{Cat}(\hat{f}^*)$. In this distribution-agnostic setting, it is impossible to compute $\operatorname{Pr}(\hat{f}_N = \hat{f}^*)$, since the absence of a relation in $\hat{f}_N$ that exists in $\hat{f}^*$ does not distinguish ``impossible'' from ``possible but rare.'' Indeed, if some $p_\ell$ is extremely small, then $\operatorname{Pr}(\hat{f}_N=\hat{f}^*)$ is negligible for feasible $N$.

Therefore, to obtain a practical guarantee, one can adopt a PAC relaxation: consider only the transitions in $\hat{f}^*$ with probabilities above a user-chosen significance level, say $\beta \in (0,1)$. We call a transition $\beta$-significant if $p_\ell \ge \beta$. Rather than certifying that \emph{every} transition has been observed, we certify that the unobserved probability mass is small. This is captured by the \emph{missing mass} $M_N$, defined as the total probability of relations not allocated to $\hat{f}_N$, i.e., $M_N = \operatorname{Pr}(\hat{f}^* \setminus \hat{f}_N)$. The missing mass is a classical object in distribution learning~\cite{berend_missing_2012}; operationally, it upper-bounds the probability that the next draw yields a previously unseen label. One estimator of $M_N$ is the Good--Turing estimate~\cite{good_turings_2000}: 
    $M_N \approx \frac{c_1}{N},$
where $c_1$ is the number of labels in $\hat{f}_N$ that appear exactly once. 

Since this estimator is not necessarily conservative, McAllester and Schapire provided an upper bound on the missing mass~\cite{mcallester_convergence_2000}. With probability at least $1-\delta$,
\begin{equation}
\label{eqn:miss-mass}
     M_N \le \frac{c_1}{N} + (2\sqrt{2} + \sqrt{3})\sqrt{\frac{\ln(3/\delta)}{N}} = \overline{M}_N.
\end{equation}
Thus, if $\overline{M}_N < \beta$, then $M_N < \beta$ holds with confidence at least $1-\delta$, i.e.,
\begin{equation}
\label{eqn:cert}
    \operatorname{Pr}( M_N < \beta) \ge \operatorname{Pr} ( M_N \le \overline{M}_N ) \ge 1-\delta, \quad \overline{M}_N < \beta.
\end{equation}

This certificate has a direct interpretation for abstraction soundness under sampling. If $M_N < \beta$, then the total probability mass of unobserved transitions is below $\beta$, which implies that every $\beta$-significant transition (every region with $p_\ell \ge \beta$) must have been observed and therefore added to $\hat{f}_N$. Equivalently, with probability at least $1-\delta$, the sampled relation $\hat{f}_N$ contains all abstract transitions that occur with probability at least $\beta$ under uniform sampling $\operatorname{Unif}(X)$.

The parameter $\beta$ therefore controls the approximation level. Choosing a smaller $\beta$ tightens the guarantee (at the cost of requiring larger $N$), and the ground-truth guarantee can even be recovered if $\beta$ falls below the minimum categorical probability. Without access to $\hat{f}^*$, however, the strongest data-driven guarantee we can certify in general is coverage of all $\beta$-significant transitions.

\smallskip
\noindent
\textbf{Running example.} We illustrate the sampling-based construction of the transition relation and its PAC-style coverage certificate on the running unicycle example. Recall the state space $X = [0.0,50.0]\times[0.0,40.0]\times[-\pi,\pi]$. We draw $N=10000$ samples $x \sim \operatorname{Unif}(X)$ and, for each sample, record the induced abstract transition $(\psi(x),\psi(f(x)))$. For example, two sampled states might be $x_1=(19.5,23.8,-0.1)$ and $x_2=(12.1,9.6,-0.73)$, which abstract to $\psi(x_1) = \hat{x}[5,5,5]$ and $\psi(x_2) = \hat{x}[4,2,3]$. Then, determine the successor abstract states:
\begin{align*}
    \psi(x'_1) & = \hat{x}[6,6,5], \quad x'_1 = f(x_1) \approx (21.5,23.6,-0.09) \\
    \psi(x'_2) &= \hat{x}[4,2,3], \quad x'_2 = f(x_2) \approx (13.6,8.3,-0.65)
\end{align*}
These two samples thus contribute the transitions $(\hat{x}[5,5,5],\sigma,\hat{x}[6,6,5])$ and $(\hat{x}[4,2,3],\sigma,\hat{x}[4,2,3])$ to the sampled relation $\hat{f}_N$. Repeat this procedure over all $10000$ samples.

To certify coverage, fix a significance threshold $\beta=0.1$ and confidence parameter $\delta=0.05$. Let $c_1$ denote the number of transition labels that appear \emph{exactly once} among the $N$ observations; suppose $c_1=12$. The McAllester--Schapire upper bound on the missing mass is then
\begin{align*}
    \overline{M}_N 
    &= \frac{c_1}{N} + (2\sqrt{2} + \sqrt{3})\sqrt{\frac{\ln(3/\delta)}{N}} \\
    &= \frac{12}{10000} + (2\sqrt{2} + \sqrt{3})\sqrt{\frac{\ln(3/0.05)}{10000}} \approx 0.0935 < \beta,
\end{align*}
which enables the PAC guarantee $ \operatorname{Pr}(M_N < 0.1) \ge 0.95$. In words, after $N=10000$ samples we can assert (with at least 95\% confidence) that the total probability mass of \emph{unobserved} transitions is below $0.1$, and hence that $\hat{f}_N$ contains every transition that occurs with probability at least $10\%$ under uniform sampling from $X$.

\section{Step 3: Purge Degenerate Behaviors}
\label{sec:step3}

\subsubsection{Types of degenerate behaviors.} Assured approximation of the ground-truth model (either through over-approximation or sampling) paired with \emph{conservative model checking} ensures that guarantees made on the abstraction transfer to the ground-truth system (Definition.~\ref{def:cons-abstraction}), i.e., $s \vDash \varphi \impliedby \hat{s} \vDash \hat{\varphi}$. However, conservative abstraction often produces \emph{degenerate behaviors} that can yield pessimistic outcomes in which the abstraction violates the lifted specification even though the concrete system satisfies it (i.e., $\hat{s} \nvDash \hat{\varphi}$ but $s \vDash \varphi$). In this section, we identify two common classes of degenerate behaviors found in conservative abstractions -- \textit{spurious transitions} and \textit{self-loop transitions} -- then discuss \emph{assured self-loop erasure} and \emph{CEGAR} to purge them from the abstraction.

\noindent
\textit{Spurious transitions.} A \emph{spurious transition} is a relation between two abstract states in $\hat{f}$ that cannot occur in the ground-truth system. Formally, it is a relation $(\hat{x}, \sigma, \hat{x}')$ such that there is no mapping from $\Psi(\hat{x})$ to $\Psi(\hat{x}')$ through $f$, i.e., the image $f(\Psi(\hat{x}))$ is disjoint from $\Psi(\hat{x}')$.

Spurious transitions arise from conservative methods for building the transition relation $\hat{f}$. These transitions appear when the set over-approximation of the image $f(\Psi(\hat{x}))$ ($\operatorname{AABB}(\cdot)$ or $\operatorname{Poly}(\cdot)$) intersects with $\Psi(\hat{x}')$, while the ground-truth image does not. These transitions are harmful in conservative model checking because they can compound into spurious \emph{counterexample trajectories} --- trajectories of the abstract system that do not satisfy $\hat{\varphi}$ and are impossible to traverse in the concrete system.

\noindent
\textit{Self-loop transitions.} A \emph{self loop} is a transition in $\hat{f}$ that maps an abstract state $\hat{x}$ to itself, formally, $(\hat{x},\sigma,\hat{x}) \in \hat{f}$. Self-loop transitions may be \emph{spurious transitions} that are artifacts of conservative successor procedures (AABB- or polytope-based methods), but they may also be an authentic representation of the ground-truth dynamics in $\Psi(\hat{x})$. The underlying issue is that the induced finite-state model is memoryless: it does not distinguish concrete states that are ``about to exit'' the cell from those that will remain inside for at least one more step. As a result, the abstraction admits the possibility of remaining in $\hat{x}$ for infinitely many steps, since a scheduler may repeatedly select the self-loop.

This is particularly harmful when reasoning about liveness specifications. Since model checking considers worst-case scheduling, it will return the infinite self-loop trace $(\hat{x}, \hat{x}, \dots)$ as a counterexample of a liveness specification, unless the trace trivially satisfies the specification (e.g., the state already lies in a goal set that must eventually be reached). For many applications, it is best to remove such self-loops by certifying that the concrete system can escape the cell in finite time. Spurious self-loop transitions can be removed through \textit{CEGAR}, while both spurious and non-spurious ones can be removed through \textit{direct certification}.

\subsubsection{Counterexample-guided abstraction refinement.} Compounding spurious transitions (spurious trajectories) are an artifact of successor over-approximation and may lead to \emph{spurious counterexamples}. Counterexample-guided abstraction refinement (CEGAR) is an iterative procedure with in-the-loop verification that purges spurious transitions in $\hat{f}$ along the counterexample trajectory by proving that such a transition cannot occur within the ground-truth system. The algorithm locally refines state cells $\hat{x} \in \hat{X}$ by splitting cells until the counterexample is refuted, or a stopping condition has been met (allowable refinements). Below, we explain the main operations in CEGAR (counterexample validation and refinement), and direct the readers to the original application of CEGAR to dynamical systems~\cite{goos_verification_2003} for further details.

A counterexample trajectory $\hat{\tau} = (\hat{x}_0, \dots, \hat{x}_M)$ from $\mathrm{Paths}_{\hat{s}}(\hat{x}_0)$ is an $M$-length trajectory output by the model checker when $\hat{\tau}  \nvDash \hat{\varphi}$. A critical step is determining whether this trajectory is spurious; that it, if for all $x_0 \in \Psi(\hat{x}_0)$, $\tau_s(x_0) \vDash \varphi$; i.e., all ground-truth trajectories originating from the subspace $\Psi( \hat{x}_0)$ do satisfy the LTL specification $\varphi$.

\begin{figure}[h]
    \centering
    \includegraphics[width=0.8\textwidth]{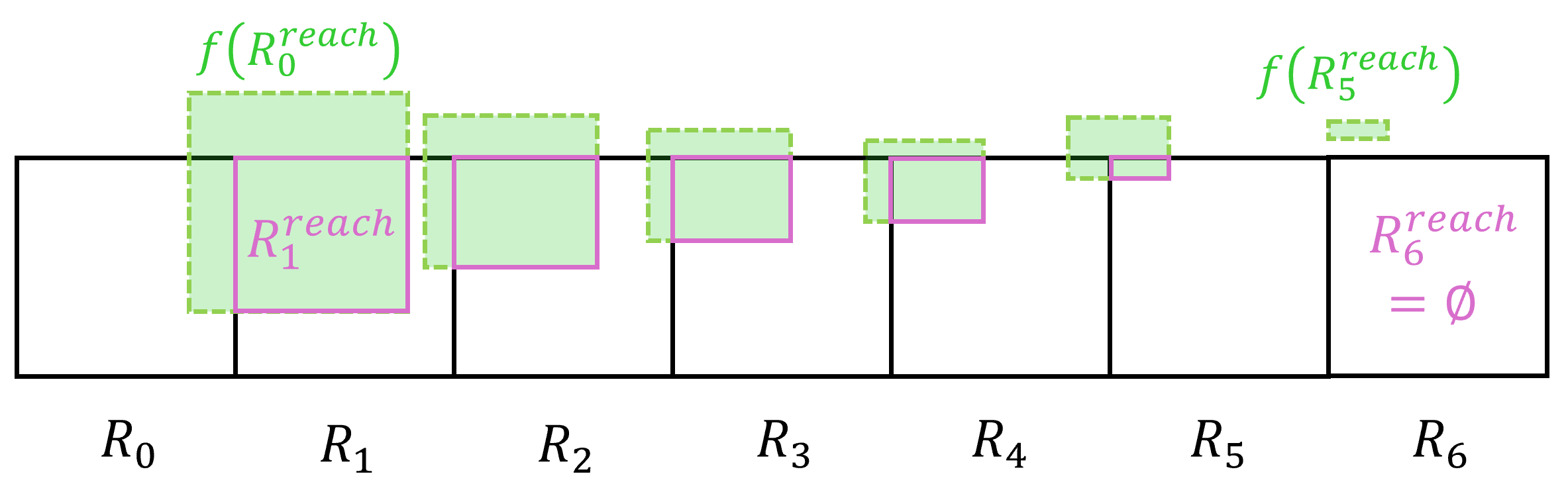}
    \vspace{-3mm}
    \caption{A counterexample trajectory admits the sequence of concrete regions $(R_0, R_1, \dots)$. Counterexample validation iteratively determines the residual intersecting area between the concretized trajectory and its abstract counterpart. In this example, this area is $0$ by $k=6$, validating that the counterexample trajectory is indeed spurious.}
    \label{fig:cegar-reach}
    \vspace{-3mm}
\end{figure}

Let $R_k = \Psi(\hat{x}_k)$ be the concretized region of $\hat{x}_k$ for $k \in \{1, \dots, M\}$. The reachable part of $R_k$ from $R_{k-1}$ is given by:
\begin{equation}
R^{reach}_0 = R_0= \Psi(\hat{x}_0), \quad R^{reach}_{k} = R_k \cap f(R^{reach}_{k-1}), \quad k=1, \dots, M
\end{equation}
If $R^{reach}_k=\emptyset$ for some $k$, then the abstract prefix cannot be realized concretely, and the counterexample trajectory is spurious. Intuitively, this recursion along $\hat{\tau} $ searches for the first instance (from the initial state) where the concrete reachable set diverges from the counterexample trajectory (see Figure~\ref{fig:cegar-reach}).

When $\hat{\tau}$ is found to be spurious, CEGAR refines the partition to block the refuted behaviors and preserve abstraction conservatism~\cite{goos_verification_2003}. Let $k$ be the first index such that $R_k=\emptyset$ in the validation recursion. Refinement then targets the transition $(\hat{x}_{k-1},\hat{x}_k)$ by \emph{splitting} $\hat{x}_k$ into a reachable subset and its complement (equivalently, splitting $\psi(\cdot)$ so these subsets become distinct abstract states) until the counterexample is refuted or the stopping condition has been met.

\subsubsection{Assured self-loop erasure.} Successor-approximation procedures may introduce self-loop transitions of the form $(\hat{x},\sigma,\hat{x})$ into $\hat{f}$. Regardless of whether these transitions are spurious or authentic, they often lead to poor verification outcomes when we attempt to verify liveness specifications through conservative model checking. It is therefore reasonable to purge a self-loop from $\hat{f}$ if we can certify a \emph{finite-time escape} property for the associated concretization. In other words, if we can show that every concrete trajectory originating in $\Psi(\hat{x})$ exits $\Psi(\hat{x})$ within $K$ steps, then we may purge the self loop. We discuss two techniques for certifying self-loop erasure: (1) an iterative bounding-box approximation of the propagated reachable set and (2) a sampling-based strategy with an associated PAC guarantee.


\smallskip
\noindent
\textit{Reachable-set propagation.} Let $R_0 = \Psi(\hat{x})$ be the initial concretized region of $\hat{x}$. Let $f(R_k)$ denote the image of $R_k$ under $f$. The propagation of $R_k$ over $k = 1, \dots, K$ is given by:
\begin{equation}
\label{eqn:reach}
    R_{k+1} = R_k \cap AABB(f(R_k)).
\end{equation}
Intuitively, $R_{k+1}$ is exactly the set of states in $R_k$ whose next-step image can also lie in $R_k$. Repeated intersection therefore filters out states that must escape, so if we find some $k \in \{1,\dots,K\}$ such that $R_k = \emptyset$, then no concrete trajectory can remain in $\Psi(\hat{x})$ for $k \le K$ steps.

\smallskip
\noindent
\textit{Sampling-based with PAC guarantee.} Suppose we impose a uniform distribution $x_0 \sim \operatorname{Unif}(\Psi(\hat{x}))$. This induces a distribution over ground truth trajectories $\tau \sim \operatorname{Pr}(\tau_s(x_0))$. Denote the $K$-length trajectory drawn from this distribution as $\tau = (x_0, \dots, x_K)$. Define the event $E = \{ \exists k \in \{1, \dots, K \} : x_k \notin \Psi(\hat{x}) \}$ to mean that the $K$-length trajectory $\tau$ has escaped $\Psi(\hat{x})$. We denote the (unknown) probability of this event as $q_{\tau} = \operatorname{Pr}(E)$.

Suppose we draw $N$ of these $K$-length trajectories $\tau \sim \operatorname{Pr}(\tau_s(x_0))$. If we observe that $E$ occurs $N$ times, then we assert the Binomial PAC guarantee:
\begin{equation}
    \operatorname{Pr}\left(q_{\tau} \ge \gamma^{1/N}\right) \ge 1-\gamma
\end{equation}
In words: with $1-\gamma$ confidence, the chance that any $K$-length trajectory escapes $\Psi(\hat{x})$ is at least $\gamma^{1/N}$. In this distribution-agnostic setting (analogous to the sample-based successor guarantee), the best we can guarantee is the coverage of successful trajectories. If we observe any $\neg E$ after $N$ trials, then we have disproven absolute coverage, and cannot remove the self loop.

\subsubsection{Running example.} We exemplify the sample-based self-loop erasure certificate on the running unicycle example. Suppose the constructed relation $\hat{f}$ contains a self-loop $(\hat{x}[5,5,5],\sigma,\hat{x}[5,5,5])$. Per Equation~\ref{eqn:bbox}, the corresponding concretization is
\begin{equation*}
    \Psi(\hat{x}[5,5,5]) = [16,20] \times [20,25] \times [-\pi/10,0].
\end{equation*}
Fix $K=10$ and a target confidence parameter $\gamma=0.1$. We draw $N=100$ initial conditions uniformly from $\Psi(\hat{x}[5,5,5])$ and simulate each trajectory for at most $K$ steps, declaring the certificate successful if every rollout exits the cell within $K$ steps. For instance, one draw yields $x_0=(18.5,22.8,-0.1)$, and propagating through the unicycle dynamics (Equation~\ref{eqn:run-dyn}) gives
\begin{equation}
    x_1=f(x_0)=
    \begin{bmatrix}
        18.5\\
        22.8\\
        -0.1
    \end{bmatrix}
    +
    \begin{bmatrix}
        2\cos(-0.1)\\
        2\sin(-0.1)\\
        0.01
    \end{bmatrix}
    \approx
    \begin{bmatrix}
        20.5\\
        22.6 \\
        -0.09
    \end{bmatrix}.
\end{equation}
which lies outside $\Psi(\hat{x}[5,5,5])$ after just one step. Observing the same behavior for all $N=100$ rollouts, we conclude the PAC-style guarantee
\begin{equation*}
    \operatorname{Pr}\!\left(p_{\tau} \ge 0.1^{1/10}\right) \ge 1-0.1 
    \;\;\equiv\;\;
    \operatorname{Pr}\!\left(p_{\tau} \ge 0.79\right) \ge 0.9.
\end{equation*}
That is, given that all $100$ sampled trajectories exit within $10$ steps, we certify with $90\%$ confidence that at least $79\%$ of trajectories starting in $\Psi(\hat{x}[5,5,5])$ also escape within $K$ steps.


\section{Step 4: Lift the LTL Specification to the Abstraction}
\label{sec:step4}

With a refined conservative abstraction $\hat{s}$ in place, the remaining task is to lift the concrete LTL specification $\varphi$ to an abstract specification $\hat{\varphi}$ over $\hat{X}$ so that conservative model checking is sound, i.e., $\hat{s} \vDash \hat{\varphi} \implies s \vDash \varphi$ (and by contraposition, $\hat{s} \nvDash \hat{\varphi} \impliedby s \nvDash \varphi$). This section introduces the additional temporal-logic machinery needed for lifting, defines a homomorphic translation $\rho$ that maps $\varphi$ to $\hat{\varphi}$, and then illustrates this lifting on the unicycle example.

\smallskip
\noindent
\textbf{May and must predicates.}
Let $AP$ denote the set of atomic propositions $p$ appearing in $\varphi$, where $p : X \rightarrow \{\top,\bot\}$. The truth set of an atom is $\sem{p} := \{ x \in X \mid p(x) = \top\}$. Given an abstract state $\hat{x} \in \hat{X}$ and its concretization $\Psi(\hat{x})$, we define \emph{may} and \emph{must} satisfaction of $p$ on $\hat{x}$ by
\begin{equation}
    \operatorname{May}_{p}(\hat{x}) \iff \Psi(\hat{x}) \cap \sem{p} \ne \emptyset,
    \qquad
    \operatorname{Must}_{p}(\hat{x}) \iff \Psi(\hat{x}) \subseteq \sem{p}.
\end{equation}
Intuitively, $p$ \emph{may} hold on $\hat{x}$ iff at least one concrete state in $\Psi(\hat{x})$ satisfies $p$, whereas $p$ \emph{must} hold on $\hat{x}$ iff every concrete state in $\Psi(\hat{x})$ satisfies $p$. We lift these to atomic predicates to the abstract state space such that:
\begin{equation}
    \hat{x} \vDash p^{\operatorname{May}} \iff \operatorname{May}_{p}(\hat{x}),
    \qquad
    \hat{x} \vDash p^{\operatorname{Must}} \iff \operatorname{Must}_{p}(\hat{x}).
\end{equation}

\noindent
\textbf{Homomorphic translation.}
Assume $\varphi$ is in negation normal form. We define a homomorphic translation operator $\rho(\cdot)$ where $\hat{\varphi} = \rho(\varphi)$. We explicitly define the operations of this function, reusing the syntax from Definition~\ref{def:ltl}:
\begin{align}
    \rho(p) &:= p^{\operatorname{Must}}, &
    \rho(\neg p) &:= \neg p^{\operatorname{May}}, \label{eq:rho-atoms} \\
    \rho(\phi_1 \wedge \phi_2) &:= \rho(\phi_1) \wedge \rho(\phi_2), &
    \rho(\phi_1 \vee \phi_2) &:= \rho(\phi_1) \vee \rho(\phi_2), \label{eq:rho-bool} \\
    \rho(\Box \phi) &:= \Box \rho(\phi), &
    \rho(\Diamond \phi) &:= \Diamond \rho(\phi), \label{eq:rho-temp1}\\
    \rho(\phi_1 ~\mathcal{U}~ \phi_2) &:= \rho(\phi_1) ~\mathcal{U}~ \rho(\phi_2), &
    \rho(\bigcirc \phi) &:= \bigcirc \rho(\phi). \label{eq:rho-temp2}
\end{align}
\looseness=-1
The temporal operators are preserved structurally, while atomic propositions are strengthened or weakened via must/may to account for state aggregation. Iterating these rules over the syntax of $\varphi$ yields a lifted specification $\hat{\varphi}$ in $\hat{X}$ which, by construction, is sound for conservative model checking, i.e., $\hat{s} \vDash \hat{\varphi} \implies s \vDash \varphi$.

\smallskip
\noindent
\textbf{Running example.} Recall the concrete LTL specification from Eqn.~\ref{eqn:run-ltl}:
\begin{equation*}
   \varphi := \bigl[(0<y_1<50)\wedge(0<y_2<40)\wedge d_1(y_1,y_2)>5\bigr] ~\mathcal{U}~ \bigl(d_2(y_1,y_2)\le 8\bigr)
\end{equation*}
For compactness, define $p_1 := (0<y_1<50)$, $p_2 := (0<y_2<40)$, $p_3 := (d_1(y_1,y_2)>5)$, and $p_4 := (d_2(y_1,y_2)\le 8)$. 
We lift the concrete specification by applying the above translation rules:
\begin{align*}
    \hat{\varphi} & = \rho(\varphi) = \rho([p_1 \wedge p_2 \wedge p_3]~\mathcal{U}~[p_4]  ), \\
    & = \rho(p_1 \wedge p_2 \wedge p_3)~\mathcal{U}~\rho(p_4) \,\,\,\, \text{(via Equation~\ref{eq:rho-temp2})}, \\
    & = \rho(p_1) \wedge \rho(p_2) \wedge \rho(p_3)~\mathcal{U}~\rho(p_4) \,\,\,\, \text{(via Equation~\ref{eq:rho-bool})}, \\
    & = \bigl[p_1^{\operatorname{Must}} \wedge p_2^{\operatorname{Must}} \wedge p_3^{\operatorname{Must}}\bigr] ~\mathcal{U}~ p_4^{\operatorname{Must}} \,\,\,\, \text{(via Equation~\ref{eq:rho-atoms})}
\end{align*}
Here $p_k^{\operatorname{Must}}$ is the \emph{must} lift of the atomic predicate $p_k$ to $\hat{X}$, defined by
\begin{equation*}
    \hat{x} \vDash p^{\operatorname{Must}}_k
    \iff \operatorname{Must}_{p_k}(\hat{x})
    \iff \Psi(\hat{x}) \subseteq \sem{p_k}.
\end{equation*}
As a concrete illustration, consider $p_1$ with truth set $\sem{p_1} := \{(y_1,y_2,\theta)\in X \mid 0<y_1<50\}$. The abstract states on which $p_1$ holds in the must sense are
\begin{equation*}
    \hat{X}_{p^{\operatorname{Must}}_1}
    := \{\hat{x}\in \hat{X} \mid \Psi(\hat{x}) \subseteq \sem{p_1}\}.
\end{equation*}
Intuitively, $\hat{X}_{p^{\operatorname{Must}}_1}$ consists of abstract states whose concretizations that do not include the boundary states. Thus, whenever $\hat{x}\in \hat{X}_{p^{\operatorname{Must}}_1}$ we have $\hat{x}\vDash p_1^{\operatorname{Must}}$, and therefore $p_1$ holds for every concrete state in $\Psi(\hat{x})$.

\section{Illustrative Case Studies}
\label{sec:exp}

The purpose of our experimental evaluation is threefold: (i) demonstrate the four-step abstraction pipeline, (ii) reveal the tradeoff between transition-building subroutine runtime and verification performance, and (iii) illustrate the improved performance after removing degenerate behaviors. For each case study and each technique, we abstract the state space, conservatively construct the transition relation via Algorithm~\ref{alg:abs-sys}, purge degenrate behaviors from the abstraction, lift the LTL specification from the concrete system to the system abstraction, and then model check the resulting abstractions against their respective LTL specifications using \texttt{pyModelChecking}~\cite{casagrande_albertocasagrandepymodelchecking_2025}.

\subsection{System Descriptions}

\subsubsection{Synthetic System.} We utilize a synthetic, linear time-invariant, discrete-time, and globally exponentially stable (GES) system with two dimensions:
\begin{equation*}
    x[k+1] = x^{*} + A(x[k] - x^{*}), \qquad A = 
    \begin{bmatrix}
        0.8 & -0.3 \\
        0.3 & 0.8
    \end{bmatrix},
    \qquad x^{*} = 
    \begin{bmatrix}
        5.0 \\
        5.0
    \end{bmatrix}
\end{equation*}
The state space of this synthetic dynamical system is given by $x = (x_1, x_2) \in X= (-10, 10) \times (-10, 10)$. The control objective is to converge to within $2.0$ units of the equilibrium point $x^{*}$ within a finite number of steps while always remaining inside of $X$. The specification is translated into LTL as:
\begin{equation}
    \varphi := \left [(-10 < x_1 < 10)\wedge (-10 < x_2 < 10)  \right] ~ \mathcal{U}\left[\|x - x^*\| \le 2.0\right]
\end{equation}

\subsubsection{Mountain Car.} We utilize the Gymnasium~\cite{towers_gymnasium_2025} MountainCar-v0 benchmark --- a widely adopted testbed for reinforcement learning --- in our evaluation. This environment models a two-dimensional control problem with state variables $x = (y, v) \in [-1.2, 0.6] \times [-0.07,0.07]$, where $y$ is the horizontal position of the vehicle and $v$ is its velocity. At each timestep, the agent selects from actions $u \in \{0, 1, 2\}$, corresponding respectively to the application of a negative force (throttle leftward), no force, or a positive force (throttle rightward).

The control objective of the Mountain Car is to reach the top of a steep incline ($y \ge 0.5$). For our experiments, we trained a lightweight deep Q-network (DQN) policy over $5000$ episodes with $\epsilon_0=1.0$ and a decay rate of $0.9983$. During closed-loop execution with non-linear dynamics, the policy is guided by ground-truth position estimates drawn from the environment at runtime. This specification can be translated into LTL as $\varphi := \Diamond (y > 0.5)$.

\smallskip
\smallskip
\noindent
\textbf{Autonomous Unicycle.} Section~\ref{sec:ovr} contains details on the unicycle case study.

\subsection{Evaluation Metrics}

\smallskip
\noindent
\textbf{Abstraction metrics.} Prior to model checking, we evaluate several descriptive metrics of the conservative abstraction $\hat{s} = (\hat{X}, \hat{X}_0, \Sigma, L, \hat{f})$. These metrics quantify \emph{degenerate behaviors} in the abstract transition system (spurious transitions and self-loop), which are often correlated with pessimistic verification outcomes. First, we report the \textit{self-loop proportion}:
\begin{equation*}
    \operatorname{SLP} := |\hat{f}_{\text{self}}|/|\hat{X}|, \quad \text{where} \quad \hat{f}_{\text{self}} := \{(\hat{x},\sigma,\hat{x}') \in \hat{f} \mid \hat{x}=\hat{x}'\}
\end{equation*}
Next, we report several successor statistics: the \textit{mean successor count}:
\begin{equation*}
    \operatorname{mSu} := \frac{1}{|\hat{X}|}\sum_{\hat{x}\in\hat{X}} |\operatorname{Post}(\hat{x})|,
\end{equation*}
the \textit{minimum and maximum successor counts}:
\begin{equation*}
    \operatorname{\underline{Su}} := \min_{\hat{x}\in\hat{X}} |\operatorname{Post}(\hat{x})| \quad \text{and} \quad  \operatorname{\overline{\text{Su}}} := \max_{\hat{x}\in\hat{X}} |\operatorname{Post}(\hat{x})|
\end{equation*}
and finally, the \textit{standard deviation of successor counts}:
\begin{equation*}
    \operatorname{Su}_{\sigma} := \sqrt{\frac{1}{|\hat{X}|}\sum_{\hat{x}\in\hat{X}}\left(|\operatorname{Post}(\hat{x})|-\operatorname{mSu}\right)^2}.
\end{equation*}


\smallskip
\noindent
\textbf{Verification metrics.} In each experiment, we treat the entire abstract state space as the set of initial states (i.e., $\hat{X}_0 = \hat{X}$) and then check the satisfaction of the LTL specification on every set of trajectories $\mathrm{Paths}_{\hat{s}}(\hat{x}_0)$ stemming from each $\hat{x}_0 \in \hat{X}_0$. This yields the satisfying subset of initial states:
\begin{equation*}
    \hat{X}_{\text{sat}} = \{\hat{x}_0 \in \hat{X}_0 \mid \mathrm{Paths}_{\hat{s}}(\hat{x}_0) \vDash \hat{\varphi}\}.
\end{equation*}
We obtain the ``ground truth'' satisfaction of $\varphi$ from the concrete system through a dense reachability analysis. Here, partition the state space $X$ with a uniform grid, inducing an abstract state space $\tilde{X}$. We denote the concretization of $\tilde{x} \in \tilde{X}$ as $\tilde{\Psi}(\tilde{x}) = \{x \mid \tilde{\psi}(x) = \tilde{x}   \}$. This reachability returns the ground-truth satisfaction set $\tilde{X}_{\text{safe}} \subseteq X$ that likely satisfy the LTL specification. This enables evaluation of the \emph{true positive rate} (TPR), which measures the proportion of truly safe states in $\hat{X}$ that were verified:
\begin{equation}
\operatorname{TPR} := \frac{\sum_{\hat{x} \in \hat{X}_{\text{sat}}} \mathbb{I}\left [ \Psi(\hat{x}) \subset\bigcup_{\tilde{x} \in \tilde{X}_{\text{safe}}} \tilde{\Psi}(\tilde{x}) \right ]}
{\sum_{\hat{x} \in \hat{X}} \mathbb{I}\left [ \Psi(\hat{x}) \subset\bigcup_{\tilde{x} \in \tilde{X}_{\text{safe}}} \tilde{\Psi}(\tilde{x}) \right ]}
\end{equation}
Finally, we report abstraction \emph{build time}, which is the runtime of all abstraction steps, and \emph{verification time}, which is the total time of model checking the final abstraction.


\subsection{Experimental Results}

\subsubsection{Effects of model size.} We performed a small ablation on all combinations of case studies and transition-building subroutines to analyze the effect of abstraction granularity on model checking performance. In particular, we vary the size of the abstract state space along each dimensions, and report the verification results in TPR. For instance, a size of $100$ abstract states along each of the $n$ state space dimensions yields $|\hat{X}|=100^n$.

\begin{figure}[h]
    \centering
    \includegraphics[width=0.6\textwidth]{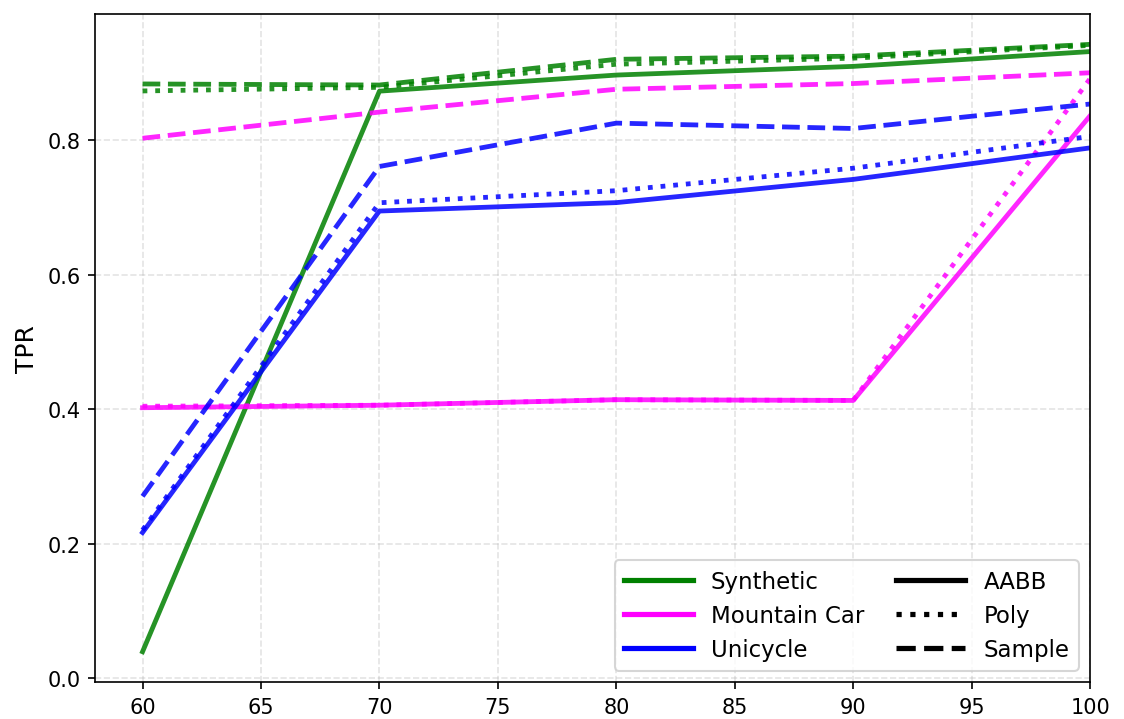}
    \vspace{-3mm}
    \caption{TPR reported against model dimension size across all case studies and transition-building subroutines.}
    \label{fig:model-size}
\end{figure}

For each case study and transition-building subroutine, we report TPR over a small window of these dimension sizes. Figure~\ref{fig:model-size} reveals that granular abstractions do indeed yield higher TPR. Furthermore, we observe the sample-based transition-building subroutine outperform its more conservative constituents.

\vspace{-4mm}

\subsubsection{Transition-building subroutines.} We analyzed the effects of each transition-building subroutine (AABB-, polytope-, and sample-based) across the case studies, reporting the model building and verification runtimes, \emph{abstraction metrics}, and \emph{verification metrics}. Abstractions were devised under fixed dimension sizes: for the synthetic case study, $|\hat{X}| = 100^2$; for mountain car, $|\hat{X}| = 100^2$; and for unicycle, $|\hat{X}| = 60^3$. We report these results under Table~\ref{tab:succ-exp}.

For the sample-based methods, we instantiate the desired certificate with confidence level $\delta = 0.01$ and significance threshold $\beta = 0.01$. Batches of 1000 samples were drawn uniformly from $\operatorname{Unif}(X)$ until the missing mass upper bound $\overline{M}_N$ (Equation~\ref{eqn:miss-mass} dropped below $\beta$, enabling us to assert the PAC guarantee in Equation~\ref{eqn:cert}). For the synthetic case study, we needed 1439 batches. For mountain car, we required 1449 batches. Finally, for unicycle, we drew 19342 batches until the significance threshold was met. Therefore, the PAC guarantee asserted for each case study is $\operatorname{Pr}(M_N < 0.01) \ge 0.99$.

Our findings reveal that less conservative transition-construction methods produce sparser transition relations (fewer self-loops and fewer total transitions). We also observe a marked reduction in the successor-count variability ($\text{Su}_\sigma$) under sampling, especially for the nonlinear systems (mountain car and unicycle). A likely explanation is that strong nonlinearities distort cell images away from rectangular shapes, causing AABB and polytope over-approximations to ``wrap'' over non-successor cells. This is not an issue in the sample-based construction because it does not admit any spurious transitions.

All together, these effects compound downstream into better verification outcomes for less conservative transition-building subroutines. We observe an increase in TPR when comparing the AABB- and polytope-based technique, and sample-based over polytope-based technique.

\begin{table}
\centering
\caption{Comparison of transition-building subroutines for each case study.}
\label{tab:succ-exp}
\setlength{\tabcolsep}{4pt}
\renewcommand{\arraystretch}{1.15}

\begin{tabular}{
  l l
  S[table-format=4.1] S[table-format=4.1]
  S[table-format=1.4] S[table-format=2.2] S[table-format=2.2] S[table-format=2.2] S[table-format=2.2]
  S[table-format=1.4]
}
\toprule
\multirow{2}{*}{Case study} &
\multirow{2}{*}{Method} &
\multicolumn{2}{c}{Time (s)} &
\multicolumn{5}{c}{Abstraction metrics} &
\multicolumn{1}{c}{Verification} \\
\cmidrule(lr){3-4}\cmidrule(lr){5-9}\cmidrule(lr){10-10}
& & {Build} & {Verify} & {SLP} & {mSu} & {\underline{Su}} & {$\overline{\text{Su}}$} & {$\text{Su}_\sigma$} & {TPR} \\
\midrule

\multirow{3}{*}{Synthetic}
  & AABB   & {0.138} & {0.235} & {0.003} & {4.07} & {1} & {9}  & {1.44} & {0.931} \\
  & PT     & {1.028} & {0.157} & {0.002} & {3.64} & {1} & {5}  & {1.12} & {0.941} \\
  & Sample & {0.782} & {0.216} & {0.002} & {3.27} & {1} & {4}  & {1.06} & {0.942} \\
\addlinespace

\multirow{3}{*}{Mountain}
  & AABB   & {0.179} & {0.158} & {0.037} & {4.64} & {1} & {80} & {3.60} & {0.836} \\
  & PT     & {2.514} & {0.261} & {0.033} & {4.37} & {1} & {51} & {2.82} & {0.891} \\
  & Sample & {2.199} & {0.144} & {0.022} & {3.77} & {1} & {10} & {0.83} & {0.900} \\
\addlinespace

\multirow{3}{*}{Unicycle}
  & AABB   & {9.130}   & {7.388} & {0.011} & {9.62} & {1} & {54} & {6.26} & {0.217} \\
  & PT     & {161.906} & {7.892} & {0.005} & {8.72} & {1} & {46} & {4.44} & {0.220} \\
  & Sample & {43.430}  & {8.238} & {0.001} & {6.68} & {1} & {22} & {1.66} & {0.271} \\
\bottomrule
\end{tabular}
\end{table}

\subsubsection{Self-loop erasure.} We examined the effects of assured self-loop erasure on the abstractions formed above (under the same dimension sizes). In particular, we report side-by-side the SLP, TPR, and SR before and after self-loop erasure. The results of reachable-set-based erasure are reported in Table~\ref{tab:sle-reach}, and the results for sample-based erasure are reported in Table~\ref{tab:sle-sample}.

For the sample-based technique, we set the desired confidence to $\gamma = 0.01$ and maximum trajectory length to $K = 50$. Sampling at least $N = 469$ trajectories yield the guarantee $\operatorname{Pr}\!\left(q_{\tau} \ge 0.99\right) \ge 0.99$. In words, with at least 99\% confidence, at least 99\% of $K$-length trajectories escape the cell. When observed that all $N$ trajectories escape, this guarantee asserts high confidence that erasing a self loop from that particular state is acceptable.

Our findings reveal an obvious reduction in SLP post-erasure across all case studies and transition-building subroutines. For the conservative AABB-based erasure, Table~\ref{tab:sle-reach} shows a nearly complete deletion of self loops in the synthetic and unicycle case study, and a significant reduction in SLP in the mountain care case study. In the sample-based regime, we observe in Table~\ref{tab:sle-sample} complete deletion of self loops in the mountain car case study. The residual SLP in the synthetic case study is likely due to the equilibrium point in the concrete system (rending it impossible to certify self-loop erasure on the state that spans this equilibrium).

\begin{table}
\centering
\caption{Verification statistics before and after \textit{reachable-set} self-loop erasure.}
\label{tab:sle-reach}
\setlength{\tabcolsep}{4.5pt} 
\renewcommand{\arraystretch}{1.15}
\begin{tabular}{
  l l
  S[table-format=4.1] S[table-format=4.1]
  S[table-format=3.2] S[table-format=3.2]
  S[table-format=3.2] S[table-format=3.2]
}
\toprule
\multirow{2}{*}{Case study} &
\multirow{2}{*}{Method} &
\multicolumn{2}{c}{Time (s)} &
\multicolumn{2}{c}{Before} &
\multicolumn{2}{c}{After} \\
\cmidrule(lr){3-4}\cmidrule(lr){5-6}\cmidrule(lr){7-8}
& & {Build} & {Verify} & {SLP} & {TPR} & {SLP} & {TPR} \\
\midrule

\multirow{3}{*}{Synthetic}
  & AABB   & {0.162} & {0.404} & {0.003} & {0.931} & {0.000} & {0.931} \\
  & PT     & {1.075} & {0.324} & {0.003} & {0.941} & {0.000} & {0.941} \\
  & Sample & {0.802} & {0.379} & {0.003} & {0.942} & {0.000} & {0.942} \\
\addlinespace

\multirow{3}{*}{Mountain}
  & AABB   & {0.256} & {0.320} & {0.037} & {0.836} & {0.004} & {0.895} \\
  & PT     & {2.534} & {0.420} & {0.033} & {0.891} & {0.004} & {0.898} \\
  & Sample & {2.216} & {0.298} & {0.022} & {0.900} & {0.001} & {0.933} \\
\addlinespace

\multirow{3}{*}{Unicycle}
  & AABB   & {42.466}  & {40.134} & {0.011} & {0.192} & {0.000} & {0.196} \\
  & PT     & {825.700} & {42.429} & {0.005} & {0.195} & {0.000} & {0.401} \\
  & Sample & {243.865} & {46.440} & {0.001} & {0.240} & {0.000} & {0.630} \\
\bottomrule
\end{tabular}
\end{table}
\vspace{-2mm}

\begin{table}
\centering
\caption{Verification statistics before and after \textit{sample-based} self-loop erasure. Confidence level is set to $\gamma$ = 0.01; sample count is set to $N$ = 469; this asserts the guarantee $\operatorname{Pr}\!\left(p_{\tau} \ge 0.99\right) \ge 0.99$ for each successfully certified erasure.}
\label{tab:sle-sample}
\setlength{\tabcolsep}{4.5pt} 
\renewcommand{\arraystretch}{1.15}

\begin{tabular}{
  l l
  S[table-format=4.1] S[table-format=4.1]
  S[table-format=3.2] S[table-format=3.2]
  S[table-format=3.2] S[table-format=3.2]
}
\toprule
\multirow{2}{*}{Case study} &
\multirow{2}{*}{Method} &
\multicolumn{2}{c}{Time (s)} &
\multicolumn{2}{c}{Before} &
\multicolumn{2}{c}{After} \\
\cmidrule(lr){3-4}\cmidrule(lr){5-6}\cmidrule(lr){7-8}
& & {Build} & {Verify} & {SLP} & {TPR} & {SLP} & {TPR} \\
\midrule

\multirow{3}{*}{Synthetic}
  & AABB   & {0.162} & {0.404} & {0.003} & {0.931} & {0.000} & {0.931} \\
  & PT     & {1.075} & {0.324} & {0.003} & {0.941} & {0.000} & {0.941} \\
  & Sample & {0.802} & {0.379} & {0.003} & {0.942} & {0.000} & {0.942} \\
\addlinespace

\multirow{3}{*}{Mountain}
  & AABB   & {0.225} & {0.364} & {0.037} & {0.836} & {0.000} & {0.895} \\
  & PT     & {2.536} & {0.467} & {0.033} & {0.891} & {0.000} & {0.898} \\
  & Sample & {2.221} & {0.346} & {0.022} & {0.900} & {0.000} & {0.933} \\
\addlinespace

\multirow{3}{*}{Unicycle}
  & AABB   & {41.905}  & {41.142} & {0.011} & {0.192} & {0.000} & {0.196} \\
  & PT     & {825.047} & {41.432} & {0.005} & {0.195} & {0.000} & {0.401} \\
  & Sample & {242.999} & {47.437} & {0.001} & {0.240} & {0.000} & {0.630} \\
\bottomrule
\end{tabular}
\end{table}
\vspace{-4mm}

These results transfer directly to downstream verification outcomes. In the mountain car case, we see marginal improvement in TPR post-erasure. In the unicycle case, we observe susbtantial imrpovement in the TPR. Finally, for the synthetic case, we do not observe any change likely because most (if not all) self loops lie near the equilibrium where concrete step sizes are small.

\subsubsection{CEGAR.} Finally, we assess the impact of CEGAR on verification outcomes across the case studies. Initial abstractions use a grid resolution of $100$ per dimension for the synthetic system ($|\hat{X}|=100^2$) and $20$ per dimension for mountain car ($|\hat{X}|=20^2$) and unicycle ($|\hat{X}|=20^3$). We report out findings in Table~\ref{tab:cegar}.

CEGAR increases model size and construction time substantially: the refined abstractions contain roughly $11500$ states (synthetic), $6500$ states (mountain car), and $15000$ states (unicycle). Nonetheless, CEGAR consistently reduces SLP, while leaving mSu largely unchanged (with only minor fluctuations).

In terms of verification performance, CEGAR yields a modest TPR improvement on the synthetic system and a pronounced improvement on mountain car (likely due to indirectly removing self-loop-driven counterexamples during refinement). In contrast, TPR is essentially unchanged for unicycle, which we attribute to its high SLP: the abstraction admits many trivial self-loop counterexamples that are difficult for CEGAR to eliminate without more targeted erasure.

\begin{table}
\centering
\caption{Verification runtime statistics before and after CEGAR. Build time is reported in minutes, while verification time is reported in seconds.} 
\label{tab:cegar}
\setlength{\tabcolsep}{4.5pt}
\renewcommand{\arraystretch}{1.15}

\begin{tabular}{
  l l
  S[table-format=4.1] S[table-format=4.1]
  S[table-format=3.2] S[table-format=3.2] S[table-format=3.2]
  S[table-format=3.2] S[table-format=3.2] S[table-format=3.2]
}
\toprule
\multirow{2}{*}{Case study} &
\multirow{2}{*}{Method} &
\multicolumn{2}{c}{Time (m, s)} &
\multicolumn{3}{c}{Before} &
\multicolumn{3}{c}{After} \\
\cmidrule(lr){3-4}\cmidrule(lr){5-7}\cmidrule(lr){8-10}
& & {Build} & {Verify}
& {SLP} & {mSu} & {TPR}
& {SLP} & {mSu} & {TPR} \\
\midrule

\multirow{2}{*}{Synthetic}
  & AABB & {113.32} & {3.64} & {0.003} & {4.07}   & {0.931} & {0.003} & {4.54} & {0.945} \\
  & PT   & {110.86} & {3.66} & {0.003} & {3.64}   & {0.941} & {0.003} & {3.95} & {0.958} \\
\addlinespace

\multirow{2}{*}{Mountain}
  & AABB & {76.90}  & {1.66}  & {0.657} & {4.26} & {0.081} & {0.197} & {5.25} & {0.430} \\
  & PT   & {148.55}  & {1.75}  & {0.643} & {4.15} & {0.081} & {0.173} & {4.62} & {0.422} \\
\addlinespace

\multirow{2}{*}{Unicycle}
  & AABB & {33.86} & {0.33} & {0.280} & {8.90} & {0.094} & {0.155} & {10.10} & {0.045} \\
  & PT   & {89.26} & {0.27} & {0.278} & {8.35} & {0.094} & {0.148} & {6.23}  & {0.093} \\
\bottomrule
\end{tabular}
\end{table}

\vspace{-2mm}
\section{Conclusion}
\label{sec:conc}

Symbolic model checking is an effective approach to verify rich temporal properties of CPS, but its guarantees only transfer to the concrete system if its abstraction is conservative. In this tutorial we explored a rigorous workflow to build such abstractions while addressing common pitfalls encountered in model building. Specifically, the discussed workflow involved covering the state space with a state abstraction, synthesizing a transition relation that conservatively approximates the concrete dynamics, purging degenerate transitions, and soundly lifting the LTL specification from the concrete to the abstract space. We validated this workflow on three case studies, illustrating the effects of each transition-building method and refinement method on downstream verification performance. Looking forward, we will explore broader CPS case studies to stress-test this workflow and extensions of these techniques to stochastic CPS.

\bibliographystyle{splncs04}
\bibliography{ref.bib}

\newpage
\appendix

\section*{Appendix}
\label{sec:app}




\subsubsection{Unicycle state controller.}
The unicycle controller used in our experiments is a smooth, deterministic state-feedback law that selects the angular velocity $u$ by steering the system toward a desired heading $\theta_d(x)$ computed from the current state. Recall the goal region centered at $y^{\mathrm{goal}} = (y_1^{\mathrm{goal}}, y_2^{\mathrm{goal}})$ with radius $r^{\mathrm{goal}}$, and the obstacle centered at $y^{\mathrm{obs}} = (y_1^{\mathrm{obs}}, y_2^{\mathrm{obs}})$ with radius $r^{\mathrm{obs}}$.

At a high level, the controller forms a planar guidance vector by combining an \emph{attractive} component that points toward the goal with a \emph{repulsive} component that pushes the system away from the obstacle. The resulting vector determines the desired heading.

Let $y = (y_1,y_2)$ denote the positional component of the state $x$, and define the displacement from the obstacle center by
\begin{equation*}
    d^{\mathrm{obs}}(x) = y - y^{\mathrm{obs}}.
\end{equation*}
Its Euclidean norm $\|d^{\mathrm{obs}}(x)\|$ is the distance from the system to the obstacle center. We then define the obstacle clearance as the distance to the obstacle boundary,
\begin{equation*}
    c(x) = \|d^{\mathrm{obs}}(x)\| - r^{\mathrm{obs}},
\end{equation*}
and map this clearance through an exponential weighting,
\begin{equation*}
    \tilde{c}(x) = e^{-\alpha c(x)},
\end{equation*}
where $\alpha \in \mathbb{R}_{>0}$ controls the rate of decay. Thus, when the system lies on the obstacle boundary ($c(x)=0$), we have $\tilde{c}(x)=1$, and this quantity decays toward $0$ as the system moves farther from the obstacle.

Using this weight, the repulsive guidance vector is defined as
\begin{equation*}
    v^{\mathrm{rep}}(x) =
    \left[\frac{\tilde{c}(x)}{\|d^{\mathrm{obs}}(x)\|^3 + \epsilon}\right] d^{\mathrm{obs}}(x),
\end{equation*}
where $\epsilon \in \mathbb{R}_{>0}$ is a small constant included to avoid singular behavior near the obstacle center. The attractive component is simply
\begin{equation*}
    v^{\mathrm{att}}(x) = y^{\mathrm{goal}} - y.
\end{equation*}
These two terms are combined as
\begin{equation*}
    v(x) = k^{\mathrm{rep}} v^{\mathrm{rep}}(x) + k^{\mathrm{att}} v^{\mathrm{att}}(x),
\end{equation*}
where $k^{\mathrm{rep}}$ and $k^{\mathrm{att}}$ are tunable repulsion and attraction gains, respectively. Intuitively, when the system is far from the obstacle, $v^{\mathrm{rep}}(x)$ is small and the heading is dominated by attraction toward the goal; near the obstacle, the repulsive term grows in influence and steers the system away from collision.

Finally, the desired heading is extracted from the planar vector $v(x)$ via
\begin{equation*}
    \theta_d(x) = \arctan(v_1(x), v_2(x)), \quad v = (v_1, v_2)
\end{equation*}
and the control input is chosen as
\begin{equation*}
    u(x) = u_{\max}\tanh\!\bigl(k_{\theta}(\theta_d(x)-\theta)\bigr),
\end{equation*}
where $u_{\max}$ sets the allowable turn rate (so that $u \in [-u_{\max},u_{\max}]$) and $k_{\theta}$ is a tunable gain on the heading error.

\subsubsection{Concrete categorical probabilities.} Fix an abstract state $\hat{x} \in \hat{X}$ and consider its concretization $\Psi(\hat{x}) \subseteq X$. Define the one-step pre-image of this concretization under $f$ as $\operatorname{Pre}(\Psi(\hat{x})) = \{ x \in X : \psi(f(x)) = \hat{x} \}$. Intuitively, $\operatorname{Pre}(\Psi(\hat{x}))$ collects exactly those concrete states that land in $\Psi(\hat{x})$ after one time step.

We next partition the state space into regions that correspond to specific abstract transitions. Concretely, for each pair $(\hat{x},\hat{x}')\in \hat{X}\times\hat{X}$, define the region of states that start in $\Psi(\hat{x})$ and map in one step into $\Psi(\hat{x}')$:
\begin{equation}
    \mathcal{R} = \{ \operatorname{Pre}(\Psi(\hat{x}')) \cap  \Psi(\hat{x}) \mid \forall \hat{x}', \hat{x} \in \hat{X}  \}.
\end{equation}
Denote an element of this collection by $R_{\ell} \in \mathcal{R}$, where $\ell \in \{1, \dots, |R|\}$. By construction, any sample $x \in R_\ell$ maps to the subset $\Psi(\hat{x}')$, and therefore witnesses the abstract transition from $\hat{x}$ to $\hat{x}'$. That is, $\psi(f(x))=\hat{x}'$ for any $x \in R_\ell$, which forms the edge $(\hat{x},\sigma,\hat{x}')$ in the sample-built transition relation $\hat{f}^*$.

Assume that samples are drawn uniformly over $X$, i.e., $x \sim \operatorname{Unif}(X)$. This induces a categorical distribution $\operatorname{Cat}(\hat{f}^*)$ over transition relations $(\psi(x), \sigma, \psi(f(x)))$ in $\hat{f}^*$ with categories $\ell =  \{1, \dots,  |\hat{f}^*|\}$. Each triple $(\psi(x), \sigma, \psi(f(x)))$ with index $\ell$ has a probability mass $q_{\ell}$ in this $\operatorname{Cat}(\hat{f}^*)$. Intuitively, $q_{\ell}$ is the chance that after a single draw $x \sim \operatorname{Unif}(X)$, the algorithm allocates the relation $(\psi(x), \sigma, \psi(f(x)))$ to $\hat{f}$.

Concretely, the probability $q_\ell$ mass of category $\ell$ is precisely:
\begin{equation}
    q_\ell= \frac{\lambda(R_{\ell})}{\lambda(X)},
\end{equation}
where $\lambda(\cdot)$ denotes Lebesgue measure (hypervolume).


\end{document}